\documentclass[letterpaper,twocolumn,10pt]{article}
\usepackage{usenix,epsfig,endnotes}

\usepackage[table,xcdraw]{xcolor}

\usepackage{graphicx}
\usepackage{multirow}
\usepackage{array}
\usepackage{subcaption}
\usepackage[labelfont=bf]{caption}

\usepackage[sort,square,comma,numbers]{natbib}

\renewcommand{\paragraph}[1]{\vspace{0.1in}\noindent\textbf{#1.}}

\makeatletter
\def\@listi{\leftmargin\leftmargini
    \parsep 1\p@ \@plus0\p@ \@minus\p@
    \topsep 2\p@   \@plus0\p@ \@minus\p@
    \itemsep1\p@ \@plus0\p@ \@minus\p@}
\let\@listI\@listi\@listi
\makeatother

\begin{document}

\title{\Large \textbf{Exploring Political Ads on News and Media Websites During the 2024 U.S. Elections}}

\author{
{\rm Emi Yoshikawa}\\
University of Washington
\and
{\rm Franziska Roesner}\\
University of Washington
} 

\date{}


\maketitle

\begin{abstract}
Building on recent work studying content in the online advertising ecosystem, including our own prior study of political ads on the web during the 2020 U.S. elections, we analyze political ad content appearing on websites leading up to and during the 2024 U.S. elections. Crawling a set of 745 news and media websites several times from three different U.S. locations (Atlanta, Seattle, and Los Angeles), we collect a dataset of over 15000 ads, including (at least) 315 political ads, and we analyze it quantitatively and qualitatively. Among our findings: a prevalence of clickbait political news ads, echoing prior work; a seemingly new emphasis (compared to 2020) on voting safety and eligibility ads, particularly in Atlanta; and non-election related political ads around the Israel-Palestine conflict, particularly in Seattle. We join prior work in calling for more oversight and transparency of political-related ads on the web. Our dataset is available at \url{https://ad-archive.cs.washington.edu}.
\end{abstract}

\section{Introduction}

Over recent election cycles, the scale and reach of online political ads has continued to grow, with spending in the 2024 U.S. elections surpassing \$10.5 billion  ---  an increase of over \$1 billion compared to the 2020 U.S. elections \cite{Election_costs_NPR}. This trend highlights not only the financial investment campaigns make in advertising, but also the ever-expanding role of digital ads in political discourse and voter engagement.

The measurable impact of political ads on voter behavior seems to be small but present.
A study conducted by the Harvard Department of Economics suggests that political ads do not significantly affect overall voter turnout but instead shape the ``partisan composition of the electorate,'' thereby impacting \textit{who} shows up vote, which can shift the election results in closely contested regions \cite{Harvard_Political_Ad_Study}. Research by Yale political scientists suggests that the effect of political ads on voter attitudes is modest, but prevalent. On a 5-point scale, political ads resulted in only a 0.05-point increase in candidate favorability~\cite{Yale_ad_impact_study}: according to Alexander Coppock, 
``the effects we demonstrated were small but detectable and could make the difference between winning and losing a close election'' \cite{Yale_article}.

In this work, we investigate the landscape of online political ads during the 202 U.S. elections, focusing particularly on the content of ads that appear on news and media websites.\footnote{We focus on web ads, rather than social media ads, due to the relative ease of data collection via a web crawler. Moreover, though Facebook makes some data available via its Ad Library, this data includes only official political ads; in our prior work, we found significant numbers of non-official political ads, e.g., political clickbait headlines~\cite{Eric_Paper}.} We build on the methodology we used in prior work to investigate the content of political web ads during the 2020 U.S. elections~\cite{Eric_Paper}.
We consider a ``political'' ad any ad with political content, whether or not the ad was run by an official political campaign committee. Our research aims to answer: 
\begin{itemize}
    \item \textbf{RQ1: Prevalence of Political Content:} How prevalent is political content in ads appearing on news and media websites leading up to the 2024 U.S. elections? 
    \item \textbf{RQ2: Types of Political Content:} What different types of political content appear in these ads? 
    \item \textbf{RQ3: Comparing Locations and Campaigns:} How does the prevalence of different types of political ads vary across viewing locations in different political regions of the U.S.? How do our results compare to prior findings during the 2020 U.S. elections?
\end{itemize}

To explore these questions, we gathered a dataset of political advertisements appearing on the web during the month leading up to and briefly following the 2024 U.S. elections. We utilized a web crawler to collect samples of ads from news and media websites between October 4th, 2024 and November 7th, 2024. Our data collection focused on three specific regions --- Atlanta, GA; Seattle, WA; and Los Angeles, CA~--- to compare how political ads varied across different geographic locations. Through a combination of qualitative and quantitative approaches, we examined the dataset holistically, compared across locations, and analyzed the ads to identify recurring themes and new trends. 
This analysis provides insight into the types of political web advertisements during the 2024 election period and their potential influence on voters across different regions and public discourse. For example, we find: 
\begin{itemize}
    \item Regional differences in political ad content between Atlanta and Seattle  ---  though we stress that these results should be interpreted cautiously, given differing dates and methodologies for data collection in the two locations. (We omit comparisons for Los Angeles due to data quality concerns.)
    \item A prevalence of clickbait-style political ads, particularly those featuring sensational headlines about political figures or election predictions (echoing prior findings~\cite{Eric_Paper}).
    \item Voter information ads, with emerging emphasis on voter eligibility and enforcement topics (a type of ad rarely seen in our prior analysis of 2020 election ads~\cite{Eric_Paper}).
    \item The use of ads to promote political views not directly related to the election.
\end{itemize}
We discuss the potential harms posed by the political ads observed and propose recommendations for regulation and future research. We also release our full dataset of ads and metadata, which can be found here: \url{https://ad-archive.cs.washington.edu}.  

\section{Background and Related Work}
\label{background_motivation}

\subsection{Historical Context}
\label{2024elec_info}
\label{2024elec_context}

The 2024 U.S. presidential election featured Kamala Harris, the Democratic nominee and former Vice President, running with Tim Walz against Donald Trump, the Republican nominee, and his running mate J.D. Vance \cite{ballotpedia_2024_presidential_election}. This election set new records in campaign advertising, with total costs exceeding \$10.53 billion, up from \$9 billion in 2020 \cite{Election_costs_NPR}. The Harris campaign spent approximately \$880 million on ads, while the Trump campaign spent approximately \$425 million \cite{nbc_total_elec2024_spend}. These figures underscore the increasing dominance of online ads in electoral campaigns, reinforcing the need for continued research to better understand their influence and implications. Our study seeks to contribute to this.

As context for 2024, the prior 2020 U.S. election saw a significant percentage of Americans doubting its legitimacy \cite{2020_election_doubt}. Despite extensive investigations disproving these claims and identifying fewer than 475 cases of voter fraud out of over 25 million votes cast in six battleground states \cite{pbs_2020_voter_fraud_fact_check}, skepticism remains widespread. A July 2023 poll found that 38\% of Americans still believed Biden did not legitimately win the presidency and among registered voters who supported Trump in 2020, 75\% continued to doubt Biden’s legitimacy \cite{2020_election_doubt}. Our study examines how these narratives and opinions are leveraged by advertisers in political campaigns and how such campaigns shape and reinforce public perceptions.

\subsection{Targeted Advertising}
\label{microtargeting}
``Microtargeting'' is defined as ``a marketing strategy that uses consumer data and demographics to identify the interests and preferences of specific individuals or small groups to send targeted advertisements that align with their interests'' \cite{microtargeting_def}. The rise of microtargeting in digital advertising, particularly for political campaigns, enables advertisers to specifically tailor their ads based on behavioral data, interests, and demographics \cite{privacy_international_profiling_microtargeting}. Political advertisers leverage these capabilities to amplify campaign messaging and influence voter turnout or preferences \cite{privacy_international_profiling_microtargeting}. 

Previous research has highlighted ethical and regulatory challenges associated with microtargeting \cite{targeting_and_adblocker, person_vs_privacy} as well as studied its impacts. For example, one study found that microtargeting political ads can be up to 70\% more effective in swaying policy support than using a generic ad designed to appeal to the entire population~\cite{microtargeting_article}.

Our research builds on these findings by analyzing 2024 election ads across three regions: Atlanta, GA; Seattle, WA; and Los Angeles, CA. These locations provide a diverse sample to investigate how regional factors, such as political party affiliation and swing-state status, influence advertising techniques and themes.

\subsection{Regulation and Transparency for Online Political Ads}
\label{regulation_transparency}
With the increase in political ads, platforms have implemented measures to enhance regulation and transparency. Some social media companies, including Twitter, TikTok, LinkedIn, and Pinterest, have banned political ads entirely \cite{american_bar_social_media}. In contrast, Meta (formerly Facebook, which also owns Instagram) and Google remain dominant players in the political advertising space, adopting distinct regulatory approaches.

Meta mandates an authorization process for ads about social issues, elections, or politics, requiring disclaimers and adherence to regional regulations \cite{facebook_ad_policy}. Political ads are also archived in the Facebook Ad Library which provides details about sponsors and targeting criteria \cite{fb_ad_library}. However, the effectiveness of this system has been criticized due to its reliance on advertisers to self-declare political ads, potentially leaving gaps for undisclosed content \cite{wired_meta_ad_policy}. Research has also show that Meta has ``spotty and inconsistent enforcement'' and that deceptive advertising networks were successfully able to issue ads on Facebook pages \cite{propublica2024_meta_badads}. In addition, researchers have raised concerns about the Ad Library’s API, citing its inadequacy in providing comprehensive data for analysis \cite{fb_ad_archive_inadequete}.\footnote{Indeed, we also initially planned to incorporate Facebook’s Ad Library data into our analysis, specifically using the Facebook/Meta Ad Library API \cite{facebook_ads_archive_api}, but obtaining access proved excessively cumbersome within the project timeline. Facebook’s process required waiting for a physical letter with a verification code (which took approximately two weeks), and despite submitting valid U.S. identification (driver's license and passport), our application was rejected three times. Ultimately, a notarized letter was required for verification, delaying access to the point where it was no longer feasible to include data from the Facebook/Meta Ad Library in our analysis.}

Similarly, Google requires political advertisers to complete a verification process, limits targeting options for election ads, and mandates disclosures about ad sponsors \cite{google_ad_policy}. However, enforcement and oversight have been criticized, with reports suggesting insufficient action against disinformation and improper ad placements \cite{google_ad_inadaquete}.

In our work, we thus (1) focus on web ads, which can be collected using a web crawler based infrastucture, and (2) consider not only explicitly categorized political ads but also those with implicit political themes that technically do not fall under Facebook and Google's policies. 

\subsection{Related Research}
\label{prior_studies}
There exists extensive research into the ad ecosystem and its implications. This research spans various domains, including computer security, privacy, and political science. Within computer science, specifically within security and privacy, studies have largely focused on the privacy-invasive mechanisms enabling ad tracking and delivery \cite{ad_info_flow, ieee_fingerprinting, journey_to_cookie}. Our work diverges by emphasizing the content of ads and the contextual targeting strategies that influence them, rather than the mechanisms behind ad delivery and tracking.

Recent studies have also examined a range of problematic ad content \cite{whatmakesabadad}, such as clickbait \cite{Zeng2020BadNC, clickbaiting_impact, clickbait_influence}, monetization from misinformation \cite{fakenewsprofit}, and malicious ads propagating malware \cite{malware}. Building on this, role of ads has been studied in the context of several political events, including the 2020 U.S. elections~\cite{2020study, Eric_Paper} and the Russian invasion of Ukraine in 2022 \cite{russia_ukraine}.
Emerging analyses of ads from the 2024 U.S. election cycle have also already been presented~\cite{2024_propublica}.

Building on this foundation, our research focuses on ads displayed across the web specifically during the month leading up to and briefly following the 2024 U.S. elections. We analyze both officially declared political ads and those with implicit political themes, addressing limitations identified in previous studies \cite{facebook_ad_study, us_political_ad_study}. 
In particular, our work is modeled after the framework we established in prior work~\cite{Eric_Paper}. We utilize our website crawler, ad scraper, site seed list, and codebook, adapting these resources to examine political advertising during the 2024 U.S. elections.

\section{Methodology}
\label{methodology}

\subsection{Ad Data Collection}
\label{ad_data_collection}
\subsubsection{Website Crawler and Ad Scraper}
\label{website_crawler}
To collect ads, we leveraged a web crawler we developed in our prior work,  as described in our 2021 study on online political advertising during the 2020 election \cite{Eric_Paper}. The crawler, built using Puppeteer, systematically scrapes advertisements from a predefined list of seed domains. The crawler detects ads via CSS selectors from EasyList (a filter list used by ad blockers) \cite{EasyList} and excludes elements smaller than 10 pixels to avoid non-advertising content. For each ad, the crawler captures a screenshot, HTML content, and the ad's parent URL. To minimize behavioral targeting, the crawler uses clean browser profiles and runs each instance within a separate Docker container \cite{docker}, minimizing any persistent tracking data across domains.

\subsubsection{Website Selection}
\label{website_selection}
We prioritized collecting advertisements from widely visited news and media websites that represent a broad range of political perspectives and the information ecosystem. As a starting point, we referenced a list of 745 news and media websites that we used in prior work studying online political advertising during the 2020 election \cite{Eric_Paper}. This list includes national newspapers, local outlets, TV stations, and digital media. This list was derived from the Tranco Top 1 Million~\cite{tranco}, leveraging the Alexa Web Information Service (now discontinued \cite{AWS_AlexaEOL}) to classify domains as news-related. The final list was truncated to 745 sites to ensure the crawler could complete its task within a single day. Selection prioritized higher-ranked sites (ranked above 5,000 by Tranco) while incorporating lower-ranked sites through stratified sampling, ensuring representation across site rankings.\footnote{Since this list was developed in 2020, we intended to construct a more current list for our study. However, the discontinuation of the Alexa Web Information Service posed challenges. Without access to a commercial website classifier, we opted to develop a basic classifier of our own.
Unfortunately, the resulting list resulted in fewer ads and lower-quality data compared to the original list. We also experimented with using the top 1,000 sites from the 2024 Tranco list (unclassified), but this approach similarly produced inferior results. Consequently, we proceeded with the previously established list of 745 news and media websites for our ad collection.}

\subsubsection{City Selection}
\label{city_selection}
Cities were selected based on regional political significance and ad traffic, informed by ad spending data from the Facebook Ad Library \cite{fb_ad_library}. The goal was to capture a range of political landscapes and ad targeting patterns:
\begin{itemize}
    \item \textbf{Atlanta, GA:} Selected to represent a key battleground state. Atlanta's role as a focal point for political campaigns offered a distinct political context compared to Seattle and Los Angeles \cite{us2024swingstates}.
    \item \textbf{Seattle, WA:} Selected due to its local political relevance and the research team's location, allowing for data collection without the need for proxy services.
    \item \textbf{Los Angeles, CA:} Selected due to its large population and political influence. Data from the Facebook Ad Archive indicated California had the highest ad spending among regions with available proxy data centers \cite{fb_ad_library}.      
\end{itemize}

\subsubsection{Crawling the Web}
\label{simulating_crawls}

We ran our crawler both before and during the election period. We conducted crawls between October 4, 2024, and November 7, 2024: in Seattle from October 4-11, in Los Angeles October 24-29, and inn in Atlanta October 31 - November 7. These datasets are available at \url{https://ad-archive.cs.washington.edu}. 

To crawl the web from Seattle, we simply ran our crawler from our university-based server, physically located in Seattle. To view the web from the perspectives of Atlanta and Los Angeles, we utilized proxies provided by Bright Data~\cite{BrightData}. For Los Angeles, Bright Data had datacenter proxies available, which we used. For Atlanta, since there were no ISP or datacenter proxies available, we used residential proxies. To enable our use of residential proxies, Bright Data reviewed and approved our use case.
To accommodate the longer resolution times associated with proxied web requests, we adjusted page load timeouts and added additional wait times before collecting ads on each site.

Unfortunately, the data we collected via the datacenter proxy from Los Angeles was low quality, e.g., including many ads with text in Hebrew. We suspect that this is because Bright Data is an Israeli company, and though the IP address was detected as being in Los Angeles, perhaps past tracking of that set of IP addresses resulted in an inference that the user speaks Hebrew. As a result, we exclude our Los Angeles dataset from most analyses, though we discuss some anecdotal results about a few U.S. election related ads that we did see there, when appropriate.

We stress that our comparisons between Seattle and Atlanta should also be interpreted with caution, due to the differing dates (early versus late October) and methods (without versus with proxy) of data collection. 
 
\subsubsection{Ethical Considerations}
\label{ethics}
We believe our research had minimal impact on internet infrastructure and the web advertising ecosystem. To minimize the impact on advertisers' budgets, we did not click on ads during the crawling process. We also conducted only a small number of crawls visiting a modest number of websites overall (13 crawls of 745 websites). More generally, website crawling is a widely accepted methodology in security and privacy research, particularly when investigating online advertisements \cite{Eric_Paper, russia_ukraine, malvertising}. Our crawler accessed only publicly available content on websites that did not involve user data, mitigating concerns regarding privacy violations. Given that the websites we examined were popular public-facing sites, we also did not anticipate substantial impacts on residential proxy IPs.

\subsection{Identifying Political Ads}
\label{ad_content_analysis}
\subsubsection{Text Extraction and Preprocessing}
\label{text_extract_preprocess}
Since advertisements typically contain text, we created a textual representation of each ad image to facilitate large-scale analysis. To extract text from each ad, we used optical character recognition (OCR), specifically using the Tesseract OCR engine \cite{TesseractOCR}. Tesseract was chosen for its accessibility as an open-source and free tool, its extensive language support (particularly useful since not all ads were in English), and its ease of integration into our research pipeline. However, Tesseract was not always able to accurately (or at all) extract text from images. 
Ads with no extracted text were filtered out of the dataset before topic-modeling. We recognize these inaccuracies may have contributed to false negatives or false positives in our analysis. We highlight this as a limitation and as an opportunity for future work, described more in depthly later on (Section \ref{limitations_future_work}).  

To prepare the extracted text for topic modeling, we applied a series of preprocessing steps. First, we removed ads where no text was found. Next, we standardized the text by converting all characters to lowercase, removing punctuation, and filtering out stop words using the Natural Language Toolkit (NLTK) stop words module \cite{NLTK}. We chose not to de-duplicate ads, as our aim was to evaluate the ad landscape as it might appear to a typical web user, who could encounter duplicate ads on the same or different sites. Additionally, upon manual review, we observed that duplicate ads often appeared on different host websites or at varying times of day, making them valuable for gaining a holistic picture of what a typical web user would experience.

\subsubsection{Identifying Political Ads}
\label{clustering_ads}

\vspace{-0.1in}
\paragraph{Topic Modeling} To identify political ads within the dataset, we applied topic modeling to group semantically similar advertisements. This approach streamlined qualitative analysis by creating clusters of related content, reducing the need to manually review the entire dataset. Initially, we experimented with multiple algorithms, including Latent Dirichlet Allocation (LDA) \cite{LDA}. However, while LDA successfully grouped keywords, it failed to capture deeper semantic themes in our dataset. We ultimately selected BERTopic \cite{bert} for its ability to generate more coherent topics by combining c-TF-IDF with clustering.

To enhance topic accuracy and coherence, we tested different topic representation models within BERTopic on our data. Two models emerged as most effective for our data:
\begin{itemize}
    \item \textbf{KeyBERTInspired} improves topic representations by considering the semantic relationships between keywords and the documents within each topic. It uses c-TF-IDF to create representative documents for each topic, then calculates the similarity between candidate keywords and the topic embeddings for more accurate keyword selection \cite{keybertinspired}.
    \item \textbf{OpenAI's GPT-4} model generates richer topic embeddings by leveraging the OpenAI API \cite{openai2023gpt4}. We integrated GPT-4 embeddings to fine-tune topic representations based on the semantic similarities across documents in each topic.
\end{itemize}
We applied BERTopic to the three location-based datasets (Atlanta, Seattle, and Los Angeles) using both representation models, producing two sets of topic clusters for each dataset (one from KeyBERTInspired and one from OpenAI).

\paragraph{Shortlisting Political Topics} Given the topics generated by each model, we manually reviewed the keywords and processed text associated with each topic to shortlist those that appeared potentially political. When determining what topics were ``potentially political,'' we adopted a broad definition of what constituted a ``potentially political'' topic. This included any theme that referenced political themes directly or could encompass sub-themes related to politics. Topics ranged from direct political references such as `voting,' `elections,' and `Trump,' to more indirect associations like `concerns in the Middle East,' `sponsored content,', `world news,' or `unbiased coverage.' This approach prioritized identifying potential political content, even at the risk of false positives, to decrease the chance of political ads being overlooked.

\paragraph{Refining Shortlisted Topics} While the shortlisted topics captured potentially political content, we found that ads flagged by only one model often indicated a higher likelihood of false positives. This occurs when one model categorizes an ad into a potentially political topic, but the other does not, suggesting the presence of unrelated or mistaken content. For instance, an ad for an insurance plan might be mistakenly grouped into a topic labeled ``plan\_trump\_president\_vote'' by one model, while another model might more accurately group it into a topic labeled ``Health Solutions and Insurance,'' indicating a likely false positive.

With this in mind, to improve precision, we filtered the dataset to retain only ads deemed potentially political by both representation models, producing unique ``potentially political'' clusters that encapsulated the set of ``potentially political ads'' to be qualitatively coded. This step ensured a higher probability of political relevance.

However, limitations in our methodology (e.g., poor OCR performance) resulted in false negatives. As a result, while exploring the dataset, we occasionally identified ads that aligned with our broad definition of ``potentially political'' but did not appear in the ads extracted from our ``potentially political'' clusters. We labeled these as manually-identified classified ads and included them in the dataset to be qualitatively coded. Although this approach introduced some inconsistency by diverging from the clustering method, we prioritized inclusivity to provide a more comprehensive view of the political ad landscape.
We did not manually review the entire dataset. 

\subsection{ Political Ad Analysis } 
\label{political_ad_analysis}
\subsubsection{Codebook Summary}
\label{sec:codebook}
To systematically analyze the content and characteristics of advertisements in our dataset, we utilized a qualitative coding approach. We based our analysis on the codebook we developed in our prior study of political advertisements during the 2020 election \cite{Eric_Paper}. The original codebook, extensively detailed in their work \cite{Eric_Paper}, provided a robust framework for classifying political ads. It featured four high-level, mutually exclusive categories:
\begin{enumerate}
    \item \textbf{Campaigns, Policy, \& Advocacy:} Ads explicitly addressing political candidates, elections, policies, or calls to action. Subcodes included dimensions such as: \textit{Election Level}, \textit{Ad Purpose} (with its own non-exclusive subcategories: \textit{Promote candidate}, \textit{Attack Ad}, \textit{Fundraising}, \textit{Petition/Polls}, \textit{Voter Info}), and \textit{Advertiser Affiliation}.
    \item \textbf{Political News and Media:} Ads promoting political news articles, videos, sources, or events. These were further divided into subcategories/subcodes: \textit{Headlines/News Articles} or \textit{News Outlets, Programs, and Events}.
    \item \textbf{Political Products:} Ads centered on selling products or services using political imagery or content. These were further divided into subcategories/subcodes: \textit{Memorabilia}, \textit{Nonpolitical Products Using Political Topics}, and \textit{Political Services}.
    \item \textbf{Malformed/Not Political}: Ads that were misclassified, occluded, or contained insufficient information for analysis. \textit{No subcodes.}
\end{enumerate}
While we preserved the high-level structure of the original codebook, we made some modifications based on our goals and data. 
Within the \textbf{Campaigns, Policy, \& Advocacy category}, the sub-code \textit{Election Level} was excluded from our analysis.
Also, for \textit{Advertiser Affiliation}, we only classified ads as \textit{Democratic}, \textit{Republican}, or \textit{Independent} if explicitly stated. All other advertisers were categorized as \textit{Other}, a designation typically applied to advocacy organizations of varying ideological affiliations. If the ad clearly reflected a corporate brand, we classified the advertiser as a \textit{Corporate Brand}.

Additionally, based on our observations and to capture themes that we did not see in 2020~\cite{Eric_Paper}, we introduced new (non-exclusive) subcategories to the subcategory \textit{Ad Purpose} (under \textbf{Campaigns, Policy, \& Advocacy}), such as:
\begin{itemize}
    \item \textit{Anti-Semitism}: Ads addressing antisemitism or promoting awareness around it.
    \item \textit{Israel-Palestine Conflict}: Ads with messaging related to this conflict.
    \item \textit{Voter Eligibility and Enforcement}: Ads promoting election security, discouraging voter fraud, or issuing warnings against non-citizens voting.
\end{itemize}

\subsubsection{Qualitative Analysis}
\label{qual_analysis}
Using the updated codebook, we qualitatively coded all of the ads in our political ads dataset (including those manually identified). 
One researcher manually reviewed each ad, identified its category, and applied the appropriate subcodes. 
We acknowledge that relying on only a single coder is not optimal, as it may introduce greater bias and errors, as we discuss further in Section \ref{limitations_future_work}.

\begin{table*}[tb]
\centering
\small
\begin{tabular}{|l|r r|r r|r r|r r|}
\hline
& \multicolumn{8}{c|}{\textbf{Potential Political Ads }} \\
\hline
& \multicolumn{2}{c|}{\textbf{Atlanta}} & \multicolumn{2}{c|}{\textbf{Seattle}} & \multicolumn{2}{c|}{\textbf{Los Angeles}} & \multicolumn{2}{c|}{\textbf{All}} \\ 
\textbf{Model} & \textbf{Count} & \textbf{\%} & \textbf{Count} & \textbf{\%} & \textbf{Count} & \textbf{\%} & \textbf{Count} & \textbf{\%} \\ \hline
\textbf{Intersection} & \textbf{135} & \textbf{2.1\%} & \textbf{131} & \textbf{1.8\%} & \textbf{21} & \textbf{1.9\%} & \textbf{287} & \textbf{1.9\%} \\ 
\hspace{10pt}KeyBERTInspired & 254 & 3.9\% & 183 & 2.5\% & 23 & 2.0\% & 460 & 3.0\% \\ 
\hspace{10pt}OpenAI GPT-4 & 312 & 4.8\% & 177 & 2.4\% & 85 & 7.5\% & 574 & 3.8\% \\ \hline
Total Ads & 6552 & 100.0\% & 7426 & 100.0\% & 1132 & 100.0\% & 15110 & 100.0\%\\ \hline
\end{tabular}
\caption{Initial Set of Potential Political Ads by Location and Model}
\label{tab:political_ads_by_model}
\end{table*}

\begin{table*}[tb]
\small
\centering
\begin{tabular}{|r|l|l|c|c|}
\hline
 & \textbf{Model} & \textbf{Topic} & \textbf{\# Ads} & \textbf{\% of Potentially Political Ads} \\ \hline
 \rowcolor{gray!20} & KeyBERT & plan\_trump\_president\_vote& & \\ 
 \rowcolor{gray!20} \multirow{-2}{*}{1} & OpenAI & Voting for President Trump's Future Plan for America& \multirow{-2}{*}{56} & \multirow{-2}{*}{41\%}\\ 
 \hline
 \multirow{2}{*}{2}& KeyBERT & newsletter\_news\_journalism\_politics& \multirow{2}{*}{18}& \multirow{2}{*}{13\%}\\ 
 & OpenAI & Political Insider Newsletter& & \\ 
 \hline
 \rowcolor{gray!20} & KeyBERT & elections\_vote\_voting\_election& & \\ 
 \rowcolor{gray!20} \multirow{-2}{*}{3} & OpenAI & Federal Voting Laws in the United States& \multirow{-2}{*}{16} & \multirow{-2}{*}{12\%}\\ 
 \hline
 \multirow{2}{*}{4}& KeyBERT & election\_voter\_voters\_voting& \multirow{2}{*}{15}& \multirow{2}{*}{11\%}\\ 
 & OpenAI & Expert Election Scenario Analysis and Investment Guidance& & \\ 
 \hline
 \rowcolor{gray!20}& KeyBERT & unbiased\_coverage\_elections\_election& & \\ 
 \rowcolor{gray!20} \multirow{-2}{*}{5} & OpenAI & Unbiased Election Coverage and Voter Empowerment& \multirow{-2}{*}{9} & \multirow{-2}{*}{7\%} \\ 
 \hline
 \multirow{2}{*}{6}& KeyBERT & virginia\_gop\_voter\_dem & \multirow{2}{*}{9}& \multirow{2}{*}{7\%}\\ 
 & OpenAI & SCOTUS Allows Virginia to Remove Noncitizens from Voter Rolls & & \\ 
 \hline
 \rowcolor{gray!20} & KeyBERT & plan\_trump\_president\_vote& & \\ 
 \rowcolor{gray!20} \multirow{-2}{*}{7} & OpenAI & Voting on Tuesday, November th & \multirow{-2}{*}{8} & \multirow{-2}{*}{6\%} \\ 
 \hline
 \multirow{2}{*}{8} 
 & KeyBERT & election\_voter\_voters\_voting& \multirow{2}{*}{3}& \multirow{2}{*}{2\%}\\ 
 & OpenAI & Unbiased Election Coverage and Voter Empowerment& & \\ 
 \hline
 \rowcolor{gray!20} & KeyBERT & plan\_trump\_president\_vote& & \\ 
 \rowcolor{gray!20} \multirow{-2}{*}{9} & OpenAI & SCOTUS Allows Virginia to Remove Noncitizens from Voter Rolls & \multirow{-2}{*}{1} & \multirow{-2}{*}{1\%} \\ 
 \hline
\multicolumn{3}{|l|}{Total Potentially Political} & 175 & 100\% \\ \hline
\end{tabular}
\caption{Atlanta's Top Potentially Political Topic Clusters}
\label{tab:atlanta_top_topics}
\end{table*}

\begin{table*}[tb]
\centering
\small
\begin{tabular}{|r|l|l|c|c|}
\hline
 & \textbf{Model} & \textbf{Topic} & \textbf{\# Ads} & \textbf{\% of Potentially Political Ads} \\ \hline
 \rowcolor{gray!20}
 & KeyBERT & lebanon\_israel\_israeli\_israelis &  &  \\ 
 \rowcolor{gray!20} \multirow{-2}{*}{1}  & OpenAI & Conflict and Ecological Concerns in the Middle East & \multirow{-2}{*}{23} & \multirow{-2}{*}{18\%} \\ 
 \hline
 \multirow{2}{*}{2} 
 & KeyBERT & democratic\_elections\_democracy\_election & \multirow{2}{*}{22} & \multirow{2}{*}{17\%} \\ 
 & OpenAI & Democratic Elections Forecast and Politics & & \\ 
 \hline
 \rowcolor{gray!20}  
 & KeyBERT & pollster\_barron\_election\_jagger &  &  \\ 
 \rowcolor{gray!20} \multirow{-2}{*}{3} & OpenAI & Stunning Election Predictions and Trump & \multirow{-2}{*}{21} & \multirow{-2}{*}{16\%} \\ 
 \hline
 \multirow{2}{*}{4} 
 & KeyBERT & harris\_donations\_donation\_walz & \multirow{2}{*}{20} & \multirow{2}{*}{15\%} \\ 
 & OpenAI & Harris Walz Team Fundraising and Donations & & \\ 
 \hline
 \rowcolor{gray!20} 
 & KeyBERT & journalism\_news\_ads\_reporting &  &  \\ 
 \rowcolor{gray!20} \multirow{-2}{*}{5} & OpenAI & Combating Fake News and Supporting Conservative Journalism & \multirow{-2}{*}{16} & \multirow{-2}{*}{12\%} \\ 
 \hline
 \multirow{2}{*}{6} 
 & KeyBERT & climate\_planet\_temperatures\_thermostat & \multirow{2}{*}{16} & \multirow{2}{*}{12\%} \\ 
 & OpenAI & Geopolitical Impact of Climate Change on Energy Management & & \\ 
 \hline
 \rowcolor{gray!20} & KeyBERT & journalism\_news\_ads\_reporting &  &  \\ 
 \rowcolor{gray!20} \multirow{-2}{*}{7} & OpenAI & News and Journalism Apps & \multirow{-2}{*}{12} & \multirow{-2}{*}{9\%} \\ 
 \hline
 \multirow{2}{*}{8} 
 & KeyBERT & harris\_donations\_donation\_walz & \multirow{2}{*}{1} & \multirow{2}{*}{1\%} \\ 
 & OpenAI & Stunning Election Predictions and Trump & & \\ 
 \hline
 \multicolumn{3}{|l|}{Total Potentially Political} & 131 & 100\% \\ \hline
\end{tabular}
\caption{Seattle's Potentially Political Topic Clusters}
\label{tab:seattle_top_topics}
\end{table*}


\begin{table*}[tb]
    \centering
    \small
    \begin{tabular}{l|rr|rr|rr|rr}
        \hline
         & \multicolumn{2}{c|}{\textbf{Atlanta, GA}} & \multicolumn{2}{c|}{\textbf{Seattle, WA}} & \multicolumn{2}{c|}{\textbf{Los Angeles, CA}} & \multicolumn{2}{c}{\textbf{Full Dataset}} \\
        \textbf{Ad Categories}
        & \textbf{Count} & \textbf{\%} & \textbf{Count} & \textbf{\%} & \textbf{Count} & \textbf{\%} &\textbf{Count} & \textbf{\%} \\ \hline
        \textbf{Political News and Media} & 49 & 30\% & 54 & 47\% & 19 & 54\% & 122 & 39\% \\ 
        \hspace{10pt}Headline/News Articles & 12 & 7\% & 21 & 18\% & 1 & 3\% & 34 & 11\%  \\ 
        \hspace{10pt}News Outlets, Programs, Events & 37 & 23\% & 33 & 29\% & 18 & 51\% & 88 & 28\% \\ \hline
        \textbf{Campaigns, Policy, \& Advocacy} & 102 & 62\% & 51 & 44\% & 16 & 46\% & 169 & 53\% \\ 
        \textit{Purpose of Ad (not mutually exclusive)} &  &  &  &  &  & & & \\ 
        \hspace{10pt}Promote Candidate or Policy & 39 & 24\% & 19 & 17\% & 7 & 20\% & 65 & 18\% \\ 
        \hspace{10pt}Attack Opposition & 7 & 4\% & 0 & 0\% & 0 & 0\% & 7 & 2\%  \\  
        \hspace{10pt}Poll, Petition, or Survey & 2 & 1\% & 10 & 9\% & 0 & 0\% & 12 & 3\% \\ 
        \hspace{10pt}Fundraising & 3 & 2\% & 17 & 15\% & 0 & 0\% & 20 & 5\% \\
        \hspace{10pt}Voter Information & 92 & 56\% & 0 & 0\% & 13 & 37\% & 105 & 28\% \\ 
        \hspace{10pt}Voter Eligibility \& Enforcement & 41 & 25\% & 0 & 0\% & 7 & 20\% & 48 & 15\% \\
        \hspace{10pt}Israel-Palestine Conflict & 0 & 0\% & 16 & 14\% & 1 & 3\% & 17 & 5\% \\
        \hspace{10pt}Antisemitism & 0 & 0\% & 5 & 4\% & 0 & 0\% & 5 & 2\% \\
        \textit{Advertiser Affiliation} &  &  &  &  &  & & & \\ 
        \hspace{10pt}Democratic Party & 7 & 4\% & 18 & 16\% & 0 & 0\% & 25 & 8\% \\ 
        \hspace{10pt}Republican Party & 52 & 31\% & 0 & 0\% & 1 & 3\% & 53 & 17\% \\ 
        \hspace{10pt}Other (Advocacy org, any ideation) & 28 & 17\% & 27 & 23\% & 6 & 17\% & 61 & 19\% \\ 
        \hspace{10pt}Independent & 4 & 2\% & 0 & 0\% & 0 & 0\% & 4 & 1\% \\ 
        \hspace{10pt}Corporate Brand & 0 & 0\% & 4 & 3\% & 1 & 3\% & 5 & 2\% \\
        \hspace{10pt}Unknown & 8 & 5\% & 2 & 2\% & 8 & 23\% & 18 & 6\% \\ \hline
        \textbf{Political Products} & 14 & 8\% & 10 & 9\% & 0 & 0\% & 24 & 8\%  \\ 
        \hspace{10pt}Political Memorabilia & 0 & 0\% & 0 & 0\% & 0 & 0\% & 0 & 0\% \\ 
        \hspace{10pt}Nonpolitical Products Using Political Topics & 14 & 8\% & 10 & 9\% & 0 & 0\% & 24 & 8\% \\ 
        \hspace{10pt}Political Services & 0 & 0\% & 0 & 0\% & 0 & 0\% & 0 & 0\% \\ \hline
        Political Ads Subtotal & 165 & 100\% & 115 & 100\% & 35 & 100\% & 315 & 100\% \\
        \hspace{10pt}Political Ads - From Clustering & 125 &  & 92 &  & 18 &  & 235 & \\ 
        \hspace{10pt}Political Ads - Manually-identified & 40 &  & 23 &  & 17 &  & 80 & \\
        \hspace{10pt}Political Ads - Malformed & 11 &  & 41 &  & 3 &  & 55 & \\ 
        Non-Political Ads Subtotal & 6,387 &  & 7,311 &  & 1,097 &  & 14,795 & \\ 
        Total & 6,552 &  & 7,426 &  & 1,132 & & 15,110 &  \\ \hline
    \end{tabular}
    \caption{Summary of the types of ads in our dataset, based on qualitative analysis. Percentages are out of political ads in that column. Note that categories are not necessarily mutually exclusive, i.e., percentages may sum to more than 100\%. We include the Los Angeles data for completeness but stress that they should not be numerically compared to Atlanta and Seattle, due to concerns about the quality of the Los Angeles dataset (see Section~\ref{simulating_crawls}).}
    \label{tab:ad_summary}
\end{table*}

\section{Results}
\label{results}

In this section, we analyze the ads in our dataset. We consider both the results of our topic modeling, which identified potentially political topic clusters, as well as our manually-identified political ads  ---  resulting in a full dataset of 315 identified politicals ads across all three crawl locations, from among 15100 ads total. We investigate the prevalence (RQ1) and the types (RQ2) of political content across our dataset, and we compare across locations and to our 2020 findings~\cite{Eric_Paper} (RQ3). We discuss RQ1 in Section~\ref{sec:rq1} and RQ2 in Section~\ref{sec:rq2}; we discuss RQ3 results inline when comparisons are relevant.

\subsection{RQ1: Prevalence of Political Content}
\label{sec:rq1}

Between October 4 and November 7, 2024, we collected 15,110 ads from three locations: Atlanta, GA; Seattle, WA; and Los Angeles, CA. From these ads, we identified 18 potentially-political clusters, totaling 290 potentially political ads, or 1.9\% of the dataset. During manual inspection and qualitative analysis of political ads, we identified 55 ads as malformed and manually identified 80 false negatives, resulting in a total of 315 political ads. Table~\ref{tab:ad_summary} shows a summary.

Table \ref{tab:political_ads_by_model} provides the topic modeling results before manual analysis. In total, 681 topics were identified across our full dataset. Of these, we categorized 41 topics as potentially political, resulting in 18 unique potentially political clusters, which corresponds to 290 ads overall. As outlined in Section~\ref{methodology}, only ads identified as potentially political by both models (i.e., their intersection) were manually coded. After qualitative analysis, the final number of political ads differed, as some were excluded or added during manual review.

\paragraph{Low Prevalence of Political Ads}
Overall, we find that the proportion of potentially political ads is low: despite being collected during the height of election season (October 4th–November 7th, 2024) \cite{usgov2024_electioninfo}, the overall proportion of potentially political ads for all locations/datasets was only 2.1\% (315/15110).

In comparison, our 2020 study reported 3.9\% of ads as potentially political \cite{Eric_Paper}. This result may reflect broader trends, such as the shift toward social media for political discourse~\cite{pew_social_media}, or it may be the result of differences in our data collection environment (e.g., the quality of proxies or improved crawler detection by ad networks).

\paragraph{Greater Proportion of Political Ads in Atlanta than Seattle}
To investigate differences across locations, we consider only Atlanta and Seattle, due to concerns about the quality our Los Angeles dataset (see Section~\ref{simulating_crawls}. We note that the prevalence of political ads in Atlanta was slightly higher than in Seattle, 2.5\% (165/6552) compared to 1.5\% (115/7426). This result is expected, given that Atlanta was a key swing state in the 2024 U.S. election and thus likely saw more political advertising.

\subsection{RQ2: Types of Political Content}
\label{sec:rq2}

To investigate more deeply the types of political content in our dataset,
we consider both our topic modeling results as well as our in-depth manual qualitative analysis.

Tables \ref{tab:atlanta_top_topics} and \ref{tab:seattle_top_topics}
shows the potentially political topic clusters that we identified as a results of our topic modeling process.\footnote{We omit the results for Los Angeles, due to dataset quality, as discussed. We did find one Spanish-language politically-relevant cluster of 21 ads there, with topics ``electorales\_arizona\_elections\_election'' and ``Voting by Mail Security in Arizona Elections'' from the KeyBERT and OpenAI models, respectively.}
Table~\ref{tab:ad_summary} shows the results of qualitatively analyzing all potentially political ads (as well as those that were identified and added manually during exploration of the dataset) using our codebook (Section~\ref{sec:codebook}).

\paragraph{Ad Content Overview} 
Overall, considering the full dataset, as shown in Table \ref{tab:ad_summary}, more than half (53\%) of the political ads were categorized as Campaign, Policy, and Advocacy ads, with Voter Information being the most common purpose (26\%). Political News and Media ads accounted for 39\% of the total ads, while Political Products made up 8\%.

\paragraph{Differences By Location}
Some differing themes emerge between Atlanta and Seattle. (Again, we refrain from drawing conclusions about Los Angeles due to concerns about the quality of the overall dataset, and we remind the reader that the Atlanta and Seattle datasets were collected in late versus early October, respectively, which may account for some differences  ---  e.g., one might expect more political ads closer to the November election.) 

For example, Voter Information was prevalent in Atlanta (56\% of political ads there), but no such ads were seen from Seattle.  
Atlanta also had a notable fraction of ads focused on voter eligibility and enforcement (25\% of political ads there). We discuss this type of content in more depth below.
In contrast, Seattle’s political ads focused on other politically-charged issues such as the Israel-Palestine Conflict (14\% of political ads there) and Antisemitism (4\%).

These impressions also appear in the raw topic modeling results: while nearly all of Atlanta’s topics mention some form of ``vote''/``voting'', voter protection, or voter information, Seattle includes no topics centered around voting, despite having election related themes in some topics (Topic 2, 3, and 8). In contrast, Seattle's top topic, ``lebanon\_israel\_israelis'' and ``Conflict and Ecological Concerns in the Middle East'', highlights a focus on non-election themes and recent global political events. Seattle's topics also showcase other non-election political issues, reflected in Topic 6: ``climate\_planet\_temperatures\_thermostat'' and ``Geopolitical Impact of Climate Change on Energy Management'.' (We note that some of these ads turned out not to be relevant to our research questions, e.g., ads for Google Nest or A/C systems, while others we classified as using political issues to advertise products or services, e.g., an ad that reads ``Climate Preparedness - As climate challenges increase, more reliable HVAC systems are vital for protecting homes''.

The topic modeling results also suggest differences in focus between political campaigns. For example, Atlanta's most prevalent topic, compromised of ``plan\_trump\_president\_vote'' and ``Voting for President Trump's Future Plan for America'', emphasizes the prominence of Trump's campaign in Atlanta. 
By contrast, Seattle's topics feature the Democratic campaign, particularly the Harris-Walz team, as evident from its second most prevalent topic, Topic 4, consisting of ``harris\_donations\_donation\_walz'' and ``Harris Walz Team Fundraising and Donations''. This focus aligns with Seattle's liberal political leaning \cite{seattletimes_seattle_liberal}. 

\paragraph{Comparison to 2020 Findings~\cite{Eric_Paper}}
Many of the our findings echo those from 2020, including the rough distributions between Political News and Media compared to Campaigns, Policy \& Advocacy compared to Political Products.
At a more granular level, the prevalence of voter eligibility and enforcement ads was a new finding; though we had hypothesized the existence of such ads in 2020, we did not observe them at the time. In another departure from our 2020 findings, no ads related to political memorabilia or political services were observed in any location.

\vspace{0.1in}
In the subsequent sections, we investigate the content of political ads more closely, highlighting several types of content that we considered particularly noteworthy or potentially problematic: misleading and clickbait political news ads, voter eligibility and enforcement ads, and ads about non-election related political issues.

\begin{figure*}[tb]
    \centering
    \captionsetup{labelfont=bf}
    \begin{subfigure}{0.4\textwidth}
        \captionsetup{labelfont=bf}
        \hspace*{2em}\includegraphics[scale=0.6]{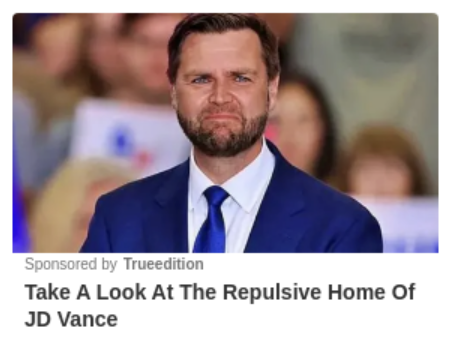}
        \subcaption{\label{fig:misleading_pol_figures_jd}}
    \end{subfigure}
    \begin{subfigure}{0.4\textwidth}
        \captionsetup{labelfont=bf}
        \hspace*{2em}\includegraphics[width=0.9\textwidth, height=1.5in, keepaspectratio]{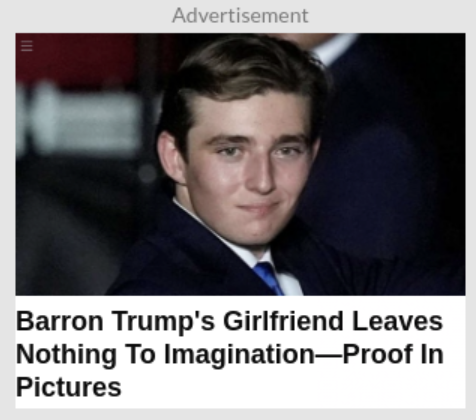}
        \subcaption{\label{fig:misleading_pol_figures_barron}}
    \end{subfigure}
    \caption{Examples of misleading political news ads utilizing sensationalized statements of political figures to entice clicks.}
    \label{fig:misleading_pol_figures}
\end{figure*}

\begin{figure*}[tb]
    \centering
    \begin{minipage}{0.3\textwidth}
       \begin{subfigure}{0.9\textwidth}
        \hspace*{2em}\includegraphics[width=0.9\textwidth, height=0.20\textheight, keepaspectratio]{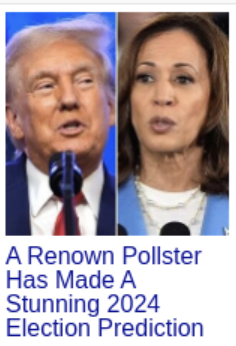}
            \caption{\label{fig:misleading_predict1}}
        \end{subfigure}
    \end{minipage}%
    \begin{minipage}{0.7\textwidth}
        \begin{subfigure}{\textwidth}
            \includegraphics[width=\textwidth, keepaspectratio]{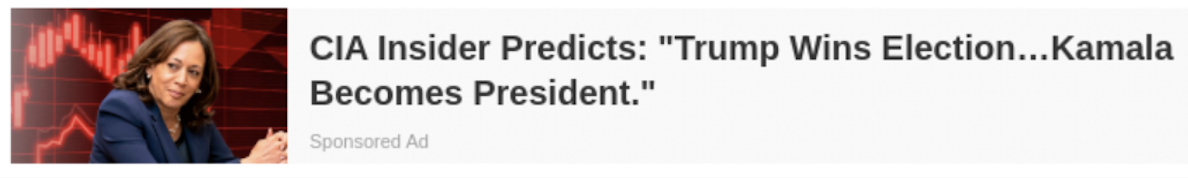}
            \caption{\label{fig:misleading_predict_CIA}}
        \end{subfigure}
        \begin{subfigure}{\textwidth}
            \includegraphics[width=\textwidth, keepaspectratio]{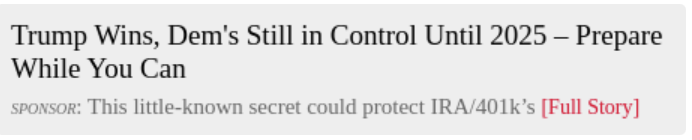}
            \caption{\label{fig:misleading_predict_demsControl}}
        \end{subfigure}
    \end{minipage}
    \caption{Examples of misleading political news ads utilizing 2024 election predictions as clickbait.}
    \label{fig:misleading_predictions}
\end{figure*}

\begin{figure*}[tb]
    \centering
    \captionsetup{labelfont=bf}
    \begin{subfigure}{0.4\textwidth}
        \captionsetup{labelfont=bf}
        \includegraphics[width=0.9\textwidth, height=2in, keepaspectratio]{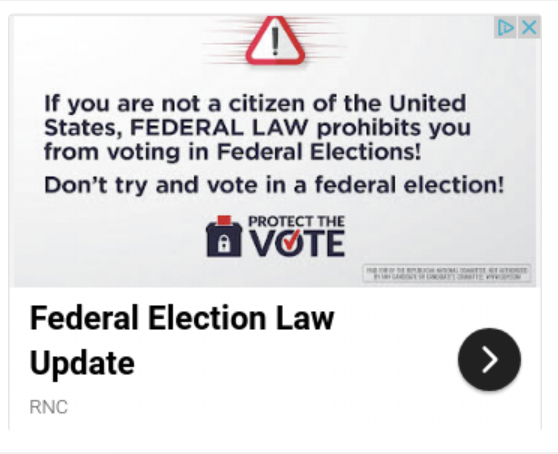}
        \caption{\label{fig:citizen_dontvote}}
    \end{subfigure}
    \begin{subfigure}{0.4\textwidth}
        \includegraphics[width=0.9\textwidth, height=2in, keepaspectratio]{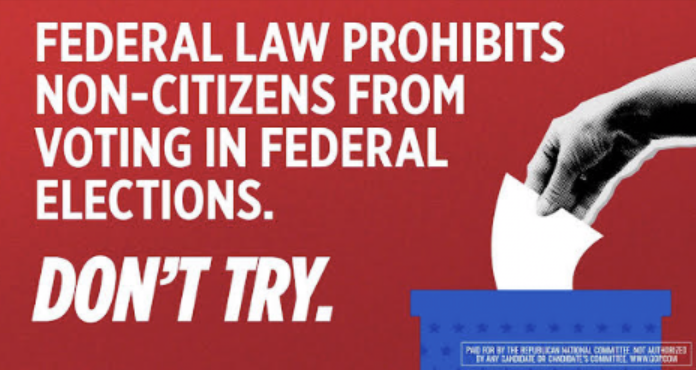}
    \caption{\label{fig:donttry}}
    \end{subfigure}
    \par\bigskip
    \begin{subfigure}{0.4\textwidth}
        \hspace*{2em}\includegraphics[width=0.9\textwidth, height=2in, keepaspectratio]{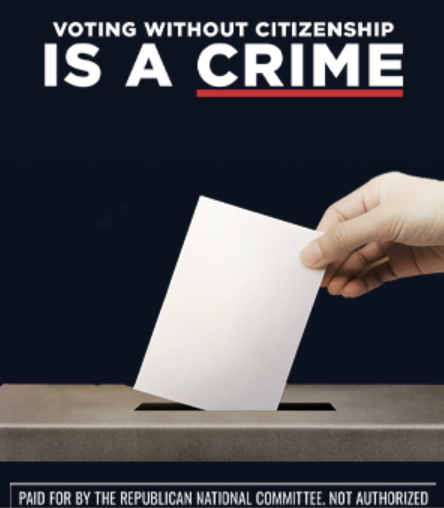}
        \subcaption[labelfont=bf]{\label{fig:crime}}
    \end{subfigure}
    \begin{subfigure}{0.4\textwidth}
        \includegraphics[width=0.9\textwidth, height=2in, keepaspectratio]{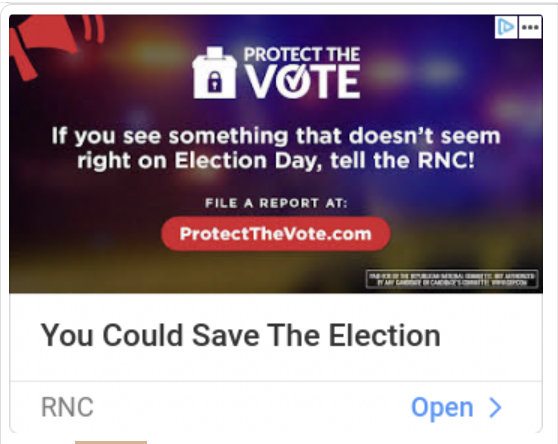}
    \caption{\label{fig:protecthevote}}
    \end{subfigure}
    \caption{Examples of ads about voter enforcement and eligibility seen from Atlanta.}
    \label{fig:voter_enforce_atlanta}
\end{figure*}

\subsubsection{Misleading Political News Ads}
\label{misleading_political_ads}
We next highlight political news and media ads that used sensationalized or exaggerated content, misleading viewers and enticing them to click. In general, advertisements often rely on strategies designed to attract attention, frequently using sensationalized headlines or misleading promises to drive clicks \cite{clickbait_influence}. We observed similar themes in our political ads dataset, specifically the Political News and Media ads, which represented 39\% of the total political ads, making them the second most prevalent type of ad in the dataset. Of these ads, 28\% were headlines/direct article links, where we observed two common themes, echoing findings from 2020~\cite{Eric_Paper}:

\paragraph{Using Political Figures for Clicks}
One recurring theme in political news ads was the use of political figures to generate interest. Ads employed misleading headlines that promised scandalous or controversial content around political individuals but redirected users to unrelated or irrelevant articles. For instance, Figure \ref{fig:misleading_pol_figures_jd} depicts an ad in which readers are urged to ``Take a look at the repulsive home of JD Vance.'' However, when followed, this link leads to a broader article about politicians' homes, without ever even mentioning Vance or his house. Similarly, the ad in Figure \ref{fig:misleading_pol_figures_barron}, makes a scandalous claim about Barron Trump’s girlfriend but directs users to a general article on athletes’ secret relationships, with no mention of Barron or his girlfriend. This trend reflects a broader strategy of using political figures to attract clicks, which has been observed in other studies and past elections~\cite{Eric_Paper}.

\paragraph{Predictions as Clickbait}
Another tactic observed was the use of predictions in political ads to drive clicks, often without providing substantial content. Ads would frequently promise predictions about elections or political events but provide little to no information. For example, Figure \ref{fig:misleading_predict1} shows an ad that promises a ``renowned pollster's 2024 election prediction'', but when followed the link leads to an article that contains no prediction, urging readers to watch a sponsored video instead. Similarly, ads like those depicted in Figures \ref{fig:misleading_predict_CIA} and \ref{fig:misleading_predict_demsControl} featured provocative headlines such as ``CIA Insider predicts Trump wins, Kamala becomes president,'' but did not substantiate the claims within the articles. These ads rely on people's curiosity about predictions, a tactic also observed in previous research into political media clickbait \cite{clickbaiting_impact}.

\subsubsection{Voter Eligibility and Enforcement}
\label{voter_e_and_e}
Compared to our 2020 findings, we observed more ads about voter information or voter eligibility and enforcement in our 2024 dataset. Overall, campaign, policy, and advocacy ads made up 53\% of the total political ads. Overall, 28\% of political ads were focused on voter information, including messages related to voter registration, election dates, voting procedures, reminders to vote, and other variations. During the 2020 election, we found only 7\% of political ads were about voter information \cite{Eric_Paper}, marking a substantial increase of such ads in 2024, at least in our datasets. 

Examining these ads more closely, we observed that many of them (46\% of political ads) were focused on voter eligibility and enforcement. These ads were primarily concerned with informing voters about eligibility requirements and encouraging vigilance regarding election integrity. The following presents key observations on the content, language, distributors, and variance across location regarding ads about voter enforcement and eligibility. 

The voter eligibility and enforcement ads in our dataset focused on reminding voters of eligibility requirements, with specific messaging targeting non-citizens. Many ads used straightforward language, such as ``Don't try and vote in a federal election!'' or ``Voting without citizenship is a crime,'' both of which can be seen in Figure \ref{fig:citizen_dontvote}. Others encouraged citizens to report suspicious behavior on election day. For example, Figure \ref{fig:protecthevote} shows an ad that reminds people, ``If you see something that doesn't seem right on Election Day, tell the RNC!'' The ad also claims ``You Could Save the Election.''

\begin{figure*}[tb]
    \centering
    \begin{subfigure}{0.4\textwidth}
        \hspace{1em}\includegraphics[width=0.9\textwidth, height=2in, keepaspectratio]{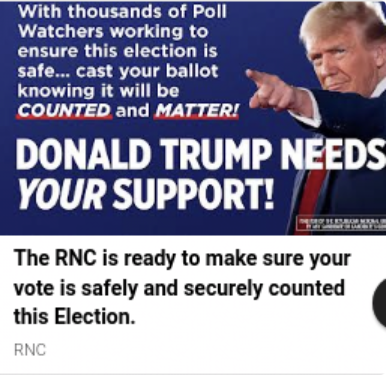}
        \caption{\label{fig:thousands_watch}}
    \end{subfigure}
    \begin{subfigure}{0.5\textwidth}
        \hspace{1em}\includegraphics[width=0.9\textwidth, height=2in, keepaspectratio]{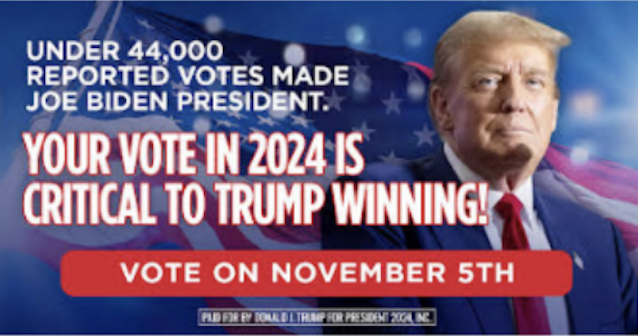}
        \caption{\label{fig:reported}}
    \end{subfigure}
    \caption{More examples of ads about voter eligibility and enforcement seen from Atlanta.}
    \label{fig:voter_enforce_atlanta_2}
\end{figure*}

Several ads emphasized vigilance, with phrases such as ``thousands of poll watchers working'' (see \ref{fig:thousands_watch}). Some ads framed their messaging by referencing past election outcomes to emphasize the importance of monitoring voting activity. For instance, Figure \ref{fig:reported} features an ad that uses the statistic, ``under 44,000 reported votes made Joe Biden President'', suggesting that the election can hinge on very few votes. 

\begin{figure*}[tb]
    \centering
    \begin{subfigure}{0.3\textwidth}
        \includegraphics[width=0.9\textwidth, height=2in, keepaspectratio]{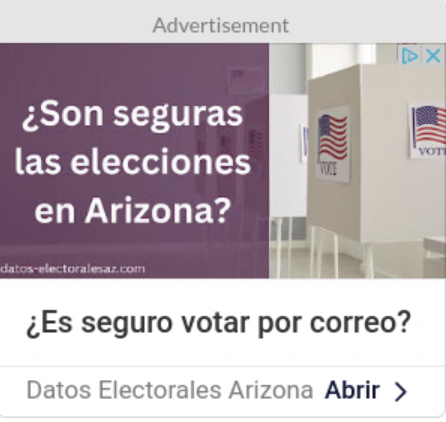}
        \caption{
            Translation: \\Are elections safe in Arizona?\\ 
            Is it safe to vote by mail?\\
            Arizona Election Data Open\>\\
        \label{fig:seguras}
        }
    \end{subfigure}
    \begin{subfigure}{0.3\textwidth}
        \includegraphics[width=0.9\textwidth, height=2in, keepaspectratio]{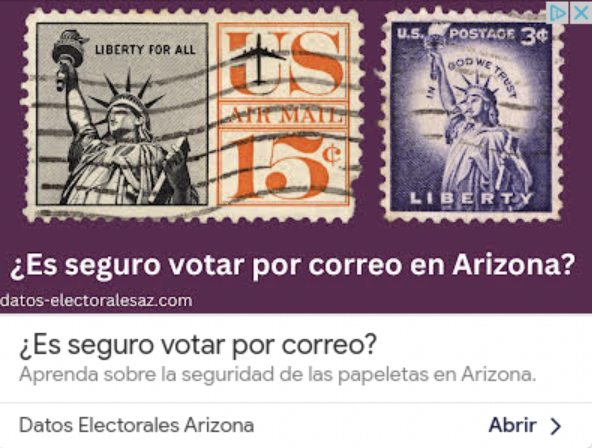}
        \caption{
            Translation: \\Is it safe to vote by mail in   \\Arizona?\\ Is it safe to vote by mail?\\
            Learn about ballot security    \\in Arizona.\\
            Arizona Election Data Open\>
        \label{fig:por_correo}
        }
    \end{subfigure}
    \begin{subfigure}{0.3\textwidth}
        \includegraphics[width=0.9\textwidth, height=2in, keepaspectratio]{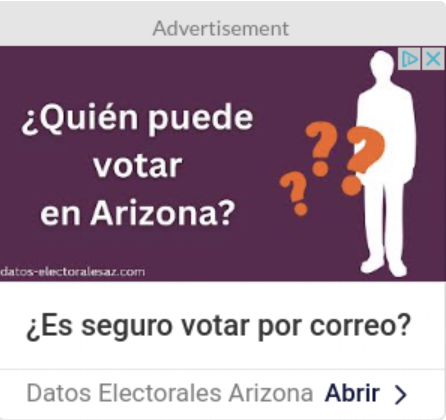}
        \caption{Translation: \\Who can vote in Arizona? \\ 
            Is it safe to vote by mail? \\
            Arizona Election Data Open\> \\
            \label{fig:quien}
        }
    \end{subfigure}
    \caption{Examples of ads about voter enforcement and eligibility seen from Los Angeles. 
    Translations by Google Translate.}
    \label{fig:voter_enforce_la}
\end{figure*}

The distribution of voter eligibility and enforcement ads exhibited distinct geographic patterns. Notably, in Atlanta, 25\% of all political ads were focused on voter eligibility and enforcement, while \textit{no} such ads were seen in Seattle. 
This difference is despite the fact that Washington has a higher percentage of foreign-born individuals than Georgia \cite{us_census_foreign_born_2018_2022}, no voter eligibility and enforcement ads were identified in Seattle. While Georgia’s status as a swing state \cite{us2024swingstates} might explain a higher concentration of ads, the extent of the disparity remained notable, warranting us to further investigate these ads and their parent URLs to identify the types of websites hosting them. Analysis revealed that many of these ads appeared on PJMedia \cite{pjmedia2024}, which is identified as a questionable source with an extreme right-wing bias by Media Bia/Fact Check \cite{mediabiasfactcheck2024}. Additional hosting sites included drudge.com \cite{drudge2024} and AL.com \cite{alcom2024}, both classified as right-center biased sources \cite{mediabiasfactcheck2024}. Based on this distribution and the absence of these ads in regions with larger foreign-born populations (e.g., California and Washington) \cite{us_census_foreign_born_2018_2022}, we hypothesize that these ads were not targeted at immigrants or non-citizens but at Republican voters and conservatives, likely aiming to stoke election-related paranoia and skepticism.

We note, anecdotally, that we also observed a cluster voter eligibility and enforcement ads in Los Angeles, though discussing voting in Arizona (perhaps due to mis-targeting). These ads are shown in Figure~\ref{fig:voter_enforce_la}.

\subsubsection{Ads About Non-Election Political Topics}
\label{non_election_ads}
While much of this study has focused on election-specific ads, a smaller fraction of the political ads in our dataset were not directly tied to electoral campaigns or 2024 election races (presidential, federal, state, etc.). This section examines how political ads were utilized as a medium for spreading and advocating views beyond the context of elections, with particular attention to advocacy related to the Israel-Palestine conflict and antisemitism.

\begin{figure*}[tb]
    \centering
    \begin{subfigure}{0.25\textwidth}
        \includegraphics[width=1.0\textwidth, height=2in, keepaspectratio]{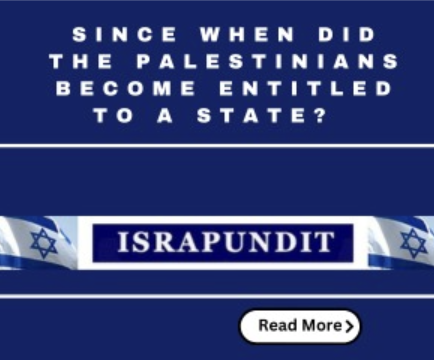}
        \caption{\label{fig:isreal_entitled}}
    \end{subfigure}
    \par\bigskip
    \begin{subfigure}{0.55\textwidth}
        \includegraphics[width=1.0\textwidth, height=3in, keepaspectratio]{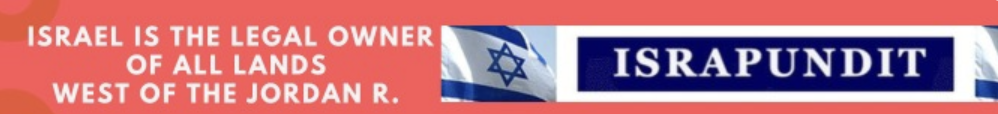}
    \caption{\label{fig:israel_legalOwner}}
    \end{subfigure}
    \caption{Examples of ads about the Israel-Palestine conflict seen from Seattle.}
    \label{fig:israel_ads}
\end{figure*}

Most of the non-election ads addressed the Israel-Palestine conflict, making up 14\% of political ads in Seattle. (No such ads appeared in Atlanta; some appeared in Los Angeles, though again we refrain from drawing conclusions due to issues with that dataset.) 
Examples of these ads can be seen in Figure \ref{fig:israel_ads}. The majority (94\%) of Israel-Palestine ads expressed pro-Israel positions, e.g., see Figure \ref{fig:israel_ads}. 
Many of these ads appeared onn websites like Israpundit, a right-leaning political commentary site that advocates for Israel's sovereignty and security \cite{israpundit2024}. 

\begin{figure*}[tb]
    \centering
    \begin{subfigure}{0.7\textwidth}
        \hspace*{2em}\vspace*{4em}\includegraphics[width=0.9\textwidth, height=2in, keepaspectratio]{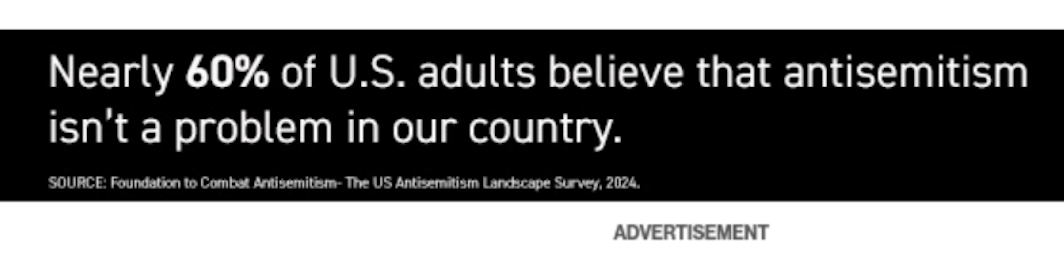}        \caption{\label{fig:antisemitism_notproblem}}
    \end{subfigure}
    \begin{subfigure}{0.2\textwidth}
        \includegraphics[width=0.9\textwidth, height=2in, keepaspectratio]{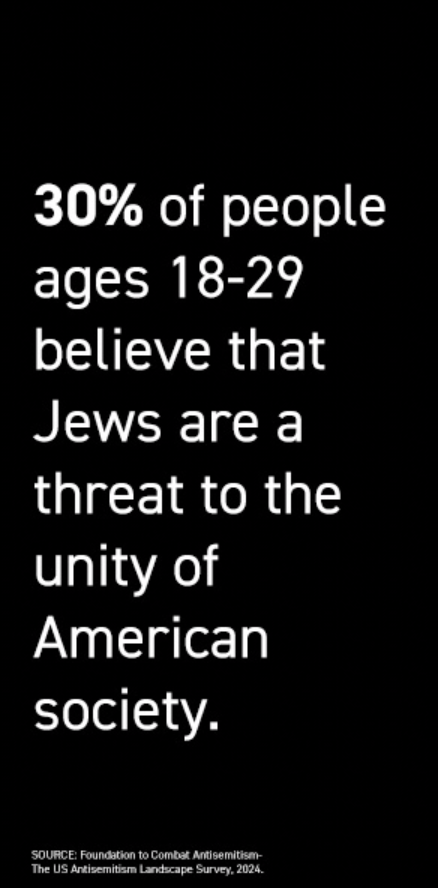}
    \caption{\label{fig:antisemitism_threat}}
    \end{subfigure}
    \caption{Examples of ads about antisemitism seen from Seattle.}
    \label{fig:antisemitism_ads}
\end{figure*}

Similarly, ads raising awareness about antisemitism appeared in 4\% of Seattle's political ads. Examples of these ads can be seen in Figure \ref{fig:antisemitism_ads}. In contrast to the Israel-Palestine conflict ads, these messages typically adopted a more neutral tone, e.g., ``Nearly 60\% of U.S. adults believe antisemitism isn’t a problem in our country.'' 

The geographic concentration of these non-election advocacy ads, particularly in Seattle, may have multiple explanations. The absence of such ads in Atlanta, a pivotal swing state, 
may suggest a strategic prioritization of election-focused messaging in competitive regions while using non-competitive areas to promote ideological or social messages. It may also simply be the result of ad inventory, with more election-related campaigns targeting Georgia as a swing state.
In any case, these campaigns demonstrate the fact that ads are used to attempt to influence public opinion on contentious and socially relevant political topics, even outside the context of elections.

\section{Discussion}

Our investigation contributes to the growing body of research on potentially problematic content in online ads, both political and otherwise (see Section \ref{prior_studies}). In this section, we first discuss the limitations of our study, highlighting  avenues for further research. We then discuss the potential harms posed by some of the political ads we identified and their implications for voter behavior and public perception. Based on those concerns, we make recommendations. 

\subsection{Limitations and Future Work}
\label{limitations_future_work}

\vspace{-0.1in}
\paragraph{Scope} First, our dataset was limited in scope, consisting of only 13 crawls over
the course of several weeks, and confined to specific geographic regions. Thus, our results
may not fully represented the broader U.S. political landscape, and there are certainly other
ad campaigns that may have run at different times, targeted at different location or segments
of the population, or that we simply did not see. Moreover, we studied only ads appearing
on websites and cannot compare our findings to political ads that may have been running
on other platforms (e.g., social media) at the same time. 
Including a more diverse set of platforms and regions in future research could provide a more holistic view of the online political advertising landscape.

Second, the study primarily analyzed the content and targeting of political ads without assessing their direct impact on viewers or public opinion. Incorporating user surveys or focus groups in future studies could yield valuable insights into the reception and influence of such ads on diverse populations. 

\paragraph{Data Collection and Methods} Data collection also presented several challenges and limitations that may have impacted the accuracy and representativeness of our results, and which could be improved in future work. The ads we collected likely do not reflect the full advertising landscape even for the selected cities, due to our limited data collection at specific times and on specific websites. We stress again that the differences in data collection dates and methods between Atlanta and Seattle may impact some comparisons, e.g., political advertising may have ramped up closer to the election during our Atlanta crawls.

Moreover, the limitations of our tools (including our crawler, OCR tool, and topic modeling) led to both false positives and false negatives in identifying politically relevant ads.
Additionally, we acknowledge the limitations of using proxies, as traffic through proxy IPs may have already been flagged by ad networks as bots or associated with unusual locations~\cite{proxy_bot}. Furthermore, our crawler does not simulate real user behavior as it lacks a realistic browsing history. As a result, our findings may not accurately reflect the ads a typical user would actually encounter.

\paragraph{Data Analysis} A further limitation lies in our qualitative coding process, which was conducted by a single researcher. While both authors discussed the codebook and potentially unclear cases in depth, the potential for subconscious bias in coding decisions cannot be eliminated entirely. 

\subsection{Concerns and Recommendations}

\vspace{-0.1in}
\paragraph{Political Clickbait}
As in our 2020 study~\cite{Eric_Paper}, we observed a significant number of attention-grabbing political ads, often designed to look like legitimate news articles, but linking to unrelated content or sources that fail to substantiate their claims (Section \ref{misleading_political_ads}). While the primary goal of these ads seems to be to generate ad revenue or drive traffic to specific websites, they pose two concerns. 
First, these ads contribute to a hyper-partisan media environment and the larger ecosystem of misinformation. They complicate the landscape of information that voters encounter, particularly when the headlines themselves lack factual backing. Second, these ads disproportionately affect more vulnerable populations, such as older adults or individuals with lower digital literacy, who may not recognize that these headlines are misleading or lead to misinformation sites. A 2020 study~\cite{stanford2020} found that 26.2\% of the population repeatedly visited websites that spread false or misleading information, with older adults being twice as likely to engage with such content. 
We argue that these types of political ads require closer scrutiny from both advertising platforms and the public. 

\paragraph{Voter Enforcement and Eligibility Ads}
Our analysis also identified ads focusing on voter eligibility and enforcement, as discussed in Section \ref{voter_e_and_e}. While these ads may be intended to inform voters of their rights and responsibilities, they often blur the line between providing useful information and inducing fear. For instance, phrases such as ``Don't try and vote!'' or ``With thousands of Poll Watchers working'' portray voting as a potentially dangerous and heavily scrutinized process. This framing may discourage voter participation by fostering concerns about eligibility challenges or legal repercussions.

The significance of these findings is heightened by the fact that many of these ads appeared on right-leaning websites, suggesting that their intended audience is Republican voters or conservatives rather than immigrant or non-citizen populations seeking information about voter rights. We hypothesize that these ads are designed to fuel election-related paranoia and polarization. This is particularly concerning as these ads target communities already vulnerable to disenfranchisement or intimidation. This approach risks further marginalizing these groups, undermining efforts to ensure equitable voter participation.
Given the potential consequences and the widespread nature of these ads, it is crucial for ad platforms and regulators to take proactive steps to monitor and address this content. 

\paragraph{Non-Election Political Ads}
In addition to election-related content, our analysis revealed a notable presence of non-election advocacy ads, specifically about the Israel-Palestine conflict and antisemitism (Section \ref{non_election_ads}). These political ads highlight the use of ad ecosystems to disseminate political content more generally, similar to our findings in prior work studying ads during the beginning of the Ukraine-Russia war~\cite{russia_ukraine}. We believe that the role of the online ad ecosystem in information dissemination around sensitive and polarizing topics such as these warrants further study and transparency. 

\paragraph{Summary of Recommendations} 
While official political ads (e.g., those run by official political campaigns) are heavily monitored and regulated, our findings suggest that other politically charged ads  ---  such as misleading clickbait or voter enforcement content  ---  continue to merit attention, even if they do not fall under current official scrutiny. We recommend that ad platforms and policymakers broaden their regulatory and transparency frameworks to include these ads, as they can still influence public opinion and contribute to misinformation. Though it is not necessarily straightforward to create policies around ad content (e.g., considering also the goals of avoiding censorship and supporting a diversity of viewpoints), as in our past work~\cite{Eric_Paper,russia_ukraine}, we continue to call for increased transparency, monitoring, and thoughtful consideration around this type of content in the ad ecosystem.

\section{Conclusion} 
This study contributes to a growing body of work examining the content and impact of political ads in the online ad ecosystem. We analyzed 15,110 ads collected from Atlanta, Seattle, and Los Angeles in the month leading up to and briefly following the 2024 U.S. elections, including 315 political ads. Using quantitative and qualitative methods, we identified regional variations in types of ads and their content, the prevalence of clickbait-style political ads, and the emergence of aggressive voter information ads, particularly those focusing on voter eligibility and enforcement. We also observed the presence of non-election political ads. We call for more rigorous scrutiny and transparency from ad platforms and regulators, alongside continued external research and monitoring of  political ads and the online ad ecosystem as a whole.

\section*{Acknowledgments}
This report is an updated version of Emi Yoshikawa's Master's thesis as part of the Paul G. Allen School of Computer Science and Engineering's BS/MS program. This work was supported in part by the National Science Foundation under Award \#2041894, by a gift from Consumer Reports, and by credits and support from BrightData.

{\bibliographystyle{acm}
\bibliography{bib}}

\end{document}